\documentstyle[12pt,a4wide]{article}
\input epsf
\renewcommand{\theequation}{\arabic{section}.\arabic{equation}}
\newcommand{\news}{\setcounter{equation}{0}}
\newcommand{\be}{\begin{equation}}
\newcommand{\ee}{\end{equation}}
\newcommand{\bea}{\begin{eqnarray}}
\newcommand{\eea}{\end{eqnarray}}
\newcommand{\bean}{\begin{eqnarray*}}
\newcommand{\eean}{\end{eqnarray*}}
\font\upright=cmu10 scaled\magstep1
\font\sans=cmss12
\newcommand{\ssf}{\sans}
\newcommand{\stroke}{\vrule height8pt width0.4pt depth-0.1pt}
\newcommand{\Z}{\hbox{\upright\rlap{\ssf Z}\kern 2.7pt {\ssf Z}}}

\newcommand{\C}{{\rlap{\rlap{C}\kern 3.8pt\stroke}\phantom{C}}}
\newcommand{\R}{\hbox{\upright\rlap{I}\kern 1.7pt R}}
\newcommand{\CP}{\C{\upright\rlap{I}\kern 1.5pt P}}
\newcommand{\half}{\frac{1}{2}}
\newcommand{\mt}{\rlap{\ssf T}\kern 3.0pt{\ssf T}}
\newcommand{\spc}{spectral curve }

\newcommand{\identity}{{\upright\rlap{1}\kern 2.0pt 1}}

\begin{document}
\pagestyle{plain}

\title{\vskip -70pt
\begin{flushright}
{\normalsize DAMTP 95-28} \\
{\normalsize To appear in Nuclear Physics B} \\
\end{flushright}
\vskip 20pt
{\bf Monopole Scattering with a Twist} \vskip 10pt
}

\author{Conor J. Houghton \\[10pt]
and \\[10pt]
Paul M. Sutcliffe\thanks{
Address from September 1995,
Institute of Mathematics,
University of Kent at Canterbury, Canterbury CT2 7NZ.
 Email P.M.Sutcliffe@ukc.ac.uk
} \\[20pt]
{\sl Department of Applied Mathematics and Theoretical Physics} \\[5pt]
{\sl University of Cambridge} \\[5pt]
{\sl Silver St., Cambridge CB3 9EW, England}\\[5pt]
{\normalsize c.j.houghton@damtp.cam.ac.uk \& p.m.sutcliffe@damtp.cam.ac.uk}
 \\[10pt]}

\date{June 1995}
\maketitle
\begin{abstract}

By imposing certain combined inversion and rotation symmetries on the
rational maps for $SU(2)$ BPS monopoles we construct geodesics in the
monopole moduli space. In the moduli space
approximation these geodesics describe a novel kind of 
monopole scattering. 
During these scattering processes axial
symmetry is instantaneously attained and, in some, monopoles
with the symmetries of the regular solids are formed.
The simplest example corresponds to a charge three monopole invariant under a 
combined inversion and $90^\circ$ rotation symmetry. In this example three well-separated
collinear unit charge monopoles coalesce
to form first a tetrahedron,
then a torus, then the dual tetrahedron and finally separate again
along the same axis of motion.
We explicitly construct the spectral curves in this case
and use a numerical ADHMN construction to compute the 
energy density at various times during the motion.
We find that the dynamics of the zeros of the Higgs field
is extremely rich and we discover a new phenomenon;
there exist charge $k$ $SU(2)$ BPS monopoles 
with more than $k$ zeros of the Higgs field.

\end{abstract}
\newpage
\section{Introduction}
\news
\ \indent 
Recently it has been shown that BPS multi-monopoles exist which
have the symmetries of the regular solids \cite{HMM,HSa,HSb}.
The subject of this paper is to investigate particularly symmetric
examples of multi-monopole scattering during which these monopole
configurations are attained. We work within the moduli space
approximation in which monopole dynamics is approximated by motion
along the geodesics in the moduli space.

The moduli space of charge $k$ multi-monopoles is equivalent to the
space of degree $k$ rational maps. In Section 2 we review this equivalence and  by
imposing twisted inversion symmetries on rational maps identify some
one-dimensional submanifolds of these maps. Since we arrive at these
submanifolds by imposing symmetry they are totally geodesic
submanifolds of the monopole moduli space and since they are one-dimensional they are geodesics. They correspond to scattering
processes in the moduli space approximation, which we refer to as twisted line scatterings.

The simplest example of a twisted line scattering involves
a charge three monopole which is symmetric under
a combined inversion and 90$^\circ$ rotation. The tetrahedral
monopole is formed twice during this 3-monopole scattering.
In Section 3 we investigate this case in more detail by constructing the
associated spectral curves and Nahm data. Using a numerical
ADHMN construction we display surfaces of constant energy density at
various times during the motion. We find that the scattering 
is particularly novel and complicated. The motion of the
zeros of the Higgs field is intriguing. There are bifurcation points
where the number of zeros of the Higgs field changes. 
This is discussed in Section 4.

\section{Monopoles and rational maps}
\news
\ \indent 
The Bogomolny equation for $SU(2)$ BPS monopoles in \R$^3$ is
\be
D_i\Phi=-\half\epsilon_{ijk}F_{jk}
\label{bog}
\ee
where $D_i=\frac{\partial}
{\partial x_i}+[A_i,$  is the covariant derivative, $A_i$ the 
{\sl su(2)}-valued gauge potential and $F_{jk}$ the gauge field. The Higgs field, $\Phi$, is an {\sl su(2)}-valued scalar
field satisfying the boundary condition
\be
\|\Phi\|=1-\frac{k}{r}+O(\frac{1}{r^2}) \hskip 10pt\mbox{as}
\hskip 10pt r\rightarrow\infty
\label{bc}
\ee
where $r=\vert\mbox{\boldmath $x$}\vert$, 
$\|\Phi\|^2=-\half\mbox{tr}\Phi^2$
and $k$ is a positive integer, known as the magnetic charge.
We shall refer to a monopole with magnetic charge $k$ as a
$k$-monopole.
The energy density, ${\cal E}$, of a monopole is given by
\be
{\cal E}=-\half\mbox{tr}(D_i\Phi)(D_i\Phi)-\frac{1}{4}
\mbox{tr}(F_{ij}F_{ij}).
\label{energy}
\ee
By substituting from the Bogomolny equation this can be rewritten in
the more convenient form 
\cite{W}, 
\be
{\cal E}=\bigtriangleup \|\Phi\|^2
\label{wardsform}.\ee
The energy is the integral of ${\cal E}$ over all space and is
equal to $8\pi k$.

It was proved by Donaldson \cite{D} that $k$-monopoles are equivalent
to rational maps \mbox{$R:\C\rightarrow\mbox{\CP}^1$}
given by
\be
R(z)=\frac{p(z)}{q(z)}
\ee
where $q$ is a monic polynomial in $z$ of degree $k$ and
$p$ is a polynomial in $z$ of degree less than $k$, which
has no factor in common with $q$. This is a 1-1 correspondence.
The rational map description of monopoles is very simple and elegant
and provides a convenient description of the monopole moduli
space. Unfortunately the rational map does not describe the detailed
properties of a monopole corresponding to
a particular point in the moduli space.

In the rational map description one
breaks the $SO(3)$ rotation symmetry of the problem
by decomposing $\R^3$ into $\R\times\C$. To do this one direction is
chosen to be special and space is decomposed into this direction and its orthogonal plane. Choosing, say, the
positive $x_3$-axis as the special direction then the $z$-coordinate
in the rational map is the complex coordinate $z=x_1+ix_2$. The only rotation
symmetries which survive in the rational map description are those
which preserve this decomposition of space.

Consider the case of a single monopole. For $k=1$ the most general
rational map of this type is
\be
R=\frac{\lambda e^{i\chi}}{z-c}
\ee
where $\lambda\in$\ \R$^+$, $\chi\in S^1, c\in$\ \C.
This describes a monopole with position $(x_1,x_2,x_3)$ and phase
angle $\chi$, where $x_1+ix_2=c$ and $x_3=\frac{1}{2}\log\lambda$.
For general $k$ there is a similar interpretation
\cite{B} of $R$ but only if the separations
of the roots of $q$ are large compared to the values of $p$ at those roots. This corresponds
to all the $k$ monopoles being well-separated in the
$x_1x_2$-plane.
In \cite{HMM} the concept of a strongly centred monopole was
introduced. Roughly, a monopole is  strongly centred if its total phase is 
unity and the centre of mass of the monopole is the origin.
In terms of the rational maps a monopole is strongly centred if and
only if the roots of $q$ sum to zero and the product of $p$ evaluated
at each of the roots of $q$ is equal to unity.
Strongly centred monopoles form a totally geodesic submanifold
of the monopole moduli space.
Here we shall be concerned only with monopoles 
which are strongly centred.

In this Section we require the following two results 
\cite{HMM} on the rational maps of cyclically symmetric and 
inversion symmetric monopoles:

\newcounter{co}
\setcounter{co}{1}
(\roman{co}) {\sl
A monopole is invariant under a rotation
by $\Theta$ around the $x_3$-axis if
\be
R(e^{i\Theta}z)\cong R(z)
\label{cycsymm}
\ee
where the equivalence means that the two rational maps are
equal up to multiplication by a phase.
}

\addtocounter{co}{1}
(\roman{co}) {\sl
Let $I$ denote the inversion operation
\be
I:(x_1,x_2,x_3)
\rightarrow
(x_1,x_2,-x_3)
\label{inv}\ee
then a monopole is inversion symmetric if}
\be
p(z)^2=1\mbox{ mod } q(z).
\ee

Note that, in general, the inversion of a monopole with rational map $p(z)/q(z)$
is the monopole with rational map $Ip(z)/q(z)$ where $Ip(z)$ is uniquely determined by
$p(z)Ip(z)=1\mbox{ mod }q(z)$.
Also, for consistency, we have followed \cite{HMM} in referring
to the transformation (\ref{inv}) as inversion. In group character tables
this is usually referred to as a reflection and denoted
$\sigma_h$.

Let $C_n$ denote the operation of rotation around the 
$x_3$-axis through an angle $2\pi/n$. It is given by
\be
C_n:(x_1,x_2,x_3)\rightarrow
(x_1\cos\frac{2\pi}{n}+x_2\sin\frac{2\pi}{n},
-x_1\sin\frac{2\pi}{n}+x_2\cos\frac{2\pi}{n},x_3).
\ee
We now generalise the inversion operation $I$ to a twisted
inversion $I_{2n}$ given by composing inversion and rotation as
\be
I_{2n}=I\circ C_{2n}.
\ee
Note that $(I_{2n})^2=C_n$ so that $I_{2n}$ symmetry implies 
$C_n$ rotation symmetry.

We now discuss charge $k$ strongly 
centred monopoles,
with $k>2$, which are
invariant under the twisted inversion symmetry $I_{2l}$, where
$l$ is an integer satisfying $k-1\geq l > k/2$.
As noted above, $I_{2l}$ symmetry implies $C_l$ symmetry,
so we first impose this condition. By  result\setcounter{co}{1}
(\roman{co}) this requires the rational map to have the form
\be
R=\frac{c+bz^l}{z^{k-l}(z^l-a)}
\ee
for some complex constants $a,b,c$.
The requirement of $I_{2l}$ symmetry for this rational map
gives the constraint
\be (c-bz^l)(c+bz^l)=1 \mbox{ mod } z^{k-l}(z^l-a).\ee
This can only be satisfied if $a=0$ and $c=\pm1$.
We can set $c=1$ by a choice of phase. We
arrive at the family of rational maps
\be
R=\frac{1+bz^l}{z^k}
\label{ratmap}
\ee
parameterized by the complex number $b$. The rational maps in this
family are strongly centred. This family defines a surface of 
two real dimensions in the $k$-monopole moduli space, which we
denote by $\Sigma_k^l$. It is a totally geodesic submanifold as it is
the fixed point set of a symmetry. $\Sigma_k^l$ is a surface of revolution;
the phase of $b$ corresponds to the orientation about the $x_3$-axis.
We may impose a reflection symmetry
on the rational map so that $b$ is real. This gives a geodesic in
$\Sigma_k^l$ corresponding to the generator of the surface of revolution.
Geodesic flow then corresponds to $b$ increasing monotonically
from $b=-\infty$ to $b=+\infty$.
If $b=0$ then (\ref{ratmap}) is the rational map of the axisymmetric
charge $k$ monopole, with the $x_3$-axis as the axis of symmetry.

In \cite{AH} pp. 25-26 it is argued that for monopoles strung out in well
separated clusters along, or nearly along, the $x_3$-axis the first
term
 in a large $z$
expansion of the rational map $R(z)$ for some phase choice will be $e^{2x}/z^L$ where $L$ is the
charge of the topmost cluster and $x$ is its elevation above the
plane. In an earlier paper \cite{HSb} we extended this and argued that if the next
highest cluster has charge $M$ and is $y$ above the plane then the
first two terms in the large $z$ expansion of the rational map will be
given by
\be R(z)\sim\frac{e^{2x}}{z^L}+\frac{e^{2y}}{z^{2L+M}}+...
\ee
Writing (\ref{ratmap}) in the form
\be R=\frac{b}{z^{k-l}}+\frac{1}{z^k} \ee
we deduce that as $b\rightarrow\pm\infty$
the rational map (\ref{ratmap}) describes axisymmetric monopoles of charge $k-l$ 
at the positions $(0,0,\pm\frac{1}{2}\log |b|)$ and
an axisymmetric charge $2l-k$ monopole at the origin.

In the moduli space approximation \cite{M,S} the dynamics
of $k$-monopoles corresponds to geodesic motion
in the $k$-monopole moduli space. Atiyah and Hitchin \cite{AH}
studied geodesics on two surfaces of revolution for the
case $k=2$. Hitchin, Manton and Murray \cite{HMM} have
investigated cyclically symmetric monopoles for $k>2$ and
obtained surfaces of revolution which are different from those
obtained here and describe a completely different type
of monopole scattering. We shall now describe the novel monopole scattering
which results from geodesic motion along the generator of $\Sigma_k^l$.

The simplest example is when $k=3$, in which case we must have $l=2$.
The rational map is
\be
R=\frac{1+bz^2}{z^3}.
\label{srat}
\ee
where $-\infty < b < \infty$. Setting $k=3$ and $l=2$ in the above
cluster decomposition we can interpret
the geodesic as the following scattering event.
At large negative times there are
three well-separated monopoles which are all located on the
$x_3$-axis. One monopole is stationary at the origin, with a second 
monopole located on the positive $x_3$-axis and a third monopole on
the negative $x_3$-axis. The second and third monopoles are
equidistant from the origin and are moving towards the 
stationary monopole. For large positive times the
situation is similar, but now the two monopoles which are on the 
positive and negative $x_3$-axis are moving away from the monopole at
the origin. We see that all the monopoles remain
on the $x_3$-axis throughout the motion, including when they
merge. This is true of all the geodesic motions along generators on
the surfaces $\Sigma_k^l$. They all describe monopole scattering along
a line. The initial configuration is of two $(k-l)$-monopoles
approaching a $(2l-k)$-monopole at the origin along the positive and
negative $x_3$-axis and the final configuration is of two
$(k-l)$-monopoles  receding along the positive and negative $x_3$-axis
leaving a $(2l-k)$-monopole at the origin. The one-parameter family of
rational maps given by (\ref{ratmap}) is invariant, up to
a phase change, under $b\rightarrow-b$ and $z\rightarrow
e^{i\pi/l}z$. This means that the outgoing configurations are always
like the incoming configurations but twisted by $\pi/l$ about the $x_3$-axis.

It is interesting that the scattering angle is zero in each case; such
zero angle scattering behaviour is normally
associated with systems which remain integrable even when time dependence
is introduced, rather than non-integrable monopole dynamics where we
expect phenomena such as right angle scattering. 

As noted above, when $b=0$ we obtain the axisymmetric $k$-monopole.
The scattering of monopoles through axisymmetric configurations is
a common occurrence which has been well studied and has analogues in
lower dimensions. However, in a twisted line scattering there is 
a subtlety in the formation of the axisymmetric monopole. 
In all previously known examples the axisymmetric $k$-monopole 
occurs in a $C_k$ symmetric monopole scattering, with the axis
of symmetry of the torus perpendicular to the plane in which 
the monopoles approach. In a twisted line scattering the axis
of symmetry of the torus is the line along which the monopoles
approach. The torus formed is perpendicular to that which one
might naively expect to find.

The $I_4$ symmetry we have imposed to obtain the surface $\Sigma_3^2$
is a combined inversion and 90$^\circ$ rotation symmetry.
A tetrahedron with vertices at $(x_1,x_2,x_3)$ given by
\be 
\{(+d,+d,+d),(+d,-d,-d),(-d,-d,+d),(-d,+d,-d)\}
\label{tetone}
\ee
has this symmetry. Here $d$ is arbitrary and the replacement
$d\rightarrow -d$ gives the dual tetrahedron.
There is a 3-monopole with tetrahedral symmetry \cite{HMM,HSa}
so for some value, $b=b_c$ say, (\ref{srat})
is the rational map of the tetrahedral monopole. When $b=-b_c$ the
tetrahedral monopole is again formed, but this time in the 
orientation dual to the previous one. 
So, although the asymptotic in and out monopole 
states may suggest
a simple scattering process, the dynamics must be relatively complicated
since a tetrahedral monopole then an axisymmetric monopole then
another tetrahedral monopole are all formed during the scattering.
In Section 3 we shall investigate in greater detail the $k=3,l=2$ 
twisted line scattering and display energy density plots which
allow us to see exactly how these various configurations are
attained during the motion.

We shall now describe 
how  the other recently discovered monopoles with the symmetries of
the regular solids may be produced in monopole scatterings. In
addition to the tetrahedral 3-monopole there is a
4-monopole with octahedral symmetry \cite{HMM} resembling a hollow
cube \cite{HSa}, a 5-monopole with octahedral symmetry resembling a
hollow octahedron and a 7-monopole with icosahedral
symmetry resembling a hollow dodecahedron \cite{HSb}.   

Orient a cube so that the $x_3$-axis goes through two opposite vertices
and the centre of the cube. This cube then has $I_6$ symmetry.
The implied $C_3$ symmetry is a rotation of the two bases of
the tetrahedra which are inscribed within the cube. The cubic
4-monopole is therefore contained within the rational maps
\be
R=\frac{1+bz^3}{z^4}
\label{ratc}
\ee
defining the surface $\Sigma_4^3$.
The cubic 4-monopole is formed twice during the 
twisted line scattering associated with the generator on $\Sigma_4^3$ where two unit charge monopoles
approach along the $x_3$-axis towards a charge two axisymmetric monopole
at the origin.
The charge four torus is also formed between the formation
of the two cubes.

There is a similar scattering through the octahedral 5-monopole. The
surface $\Sigma_5^3$ also describes monopoles with $I_6$
symmetry. An octahedron orientated so that two of its triangular faces
are parallel to the $x_1x_2$-plane is invariant under $I_6$. Thus
geodesic motion along the generator of $\Sigma_5^3$ describes two 2-monopoles approaching from the positive and
negative $x_3$-axis a single monopole at the origin; the monopoles then
coalesce to form the octahedral 5-monopole which deforms further into the toroidal 5-monopole and
then into the octahedral 5-monopole rotated through $\pi/3$ before separating
again into two 2-monopoles, receding along the $x_3$-axis and
leaving a single monopole at the origin.

A dodecahedron with two faces parallel to the $x_1x_2$-plane has
$I_{10}$ symmetry. The dodecahedral 7-monopole occurs during the
geodesic scattering on the surface $\Sigma_7^5$ of rational maps
\be
R=\frac{1+bz^5}{z^7}.
\label{rat}
\ee
The scattering involves two axisymmetric 2-monopoles approaching from
the positive and negative $x_3$-axis a  axisymmetric
charge three monopole at the origin. The
dodecahedral monopole is formed, followed by the axisymmetric charge
seven monopole, then the dodecahedral monopole rotated $\pi/5$ relative
to the previous one. Finally two 2-monopoles separate out again along
the $x_3$-axis leaving a charge three monopole behind.

Scattering geodesics which include the tetrahedral and cubic
monopoles have been obtained previously \cite{HMM} by imposing
cyclic $C_3$ and dihedral $D_4$ symmetries respectively. 
A scattering geodesic containing the octahedral 5-monopole also
exists \cite{HSb} resulting from imposition of $D_4$ symmetry.
A further scattering geodesic through the cubic monopole can
be obtained by imposition of tetrahedral symmetry on four monopoles
\cite{HSa}. All these examples are very different from the twisted
line scatterings which we present here. Furthermore, only the last
example, that of 4-monopole scattering with tetrahedral symmetry, has
been studied in great detail with the evolution of the energy
density examined. The twisted line scatterings appear to
be more complicated than any of those above and one
really needs to examine energy density surfaces to understand
these processes.

\section{The charge three twisted line scattering}
\news\ \indent
We now investigate the 3-monopoles along the $\Sigma_3^2$ geodesic in
more detail in order to plot their energy densities. We
use constructions
similar to those in \cite{HMM,HSa,HSb} and these papers should
be consulted for a more detailed explanation. We
exploit two different formulations of the monopole problem; spectral
curves and Nahm data.

Monopoles correspond to certain algebraic curves, called spectral
curves, in the tangent bundle to the Riemann sphere T\CP$^1$. We
let $\zeta$ be the standard inhomogeneous coordinate on the base space
and $\eta$ the fibre coordinate. 
Monopoles of charge $k$ correspond to curves of the form
\be
\eta^k+\eta^{k-1} a_1(\zeta)+\ldots+\eta^r a_{k-r}(\zeta)+
\ldots+\eta a_{k-1}(\zeta)+a_k(\zeta)=0.\label{algcurve}
\label{gensc}
\ee
where, for $1\leq r\leq k$, $a_r(\zeta)$ is a polynomial in $\zeta$ of
maximum degree $2r$ and satisfies the reality condition
\be 
a_r(\zeta)=(-1)^r\zeta^{2r}\overline{a_r(-\frac{1}{\overline{\zeta}})}
.\ee

A general algebraic curve in T\CP$^1$ satisfying these conditions will not
necessarily correspond to a monopole. It is difficult to demonstrate a
particular algebraic curve is the spectral curve of a monopole. We
will circumvent these difficulties by using the ADHMN
construction. Before we do so we will construct a candidate algebraic
curve for $I_4$ symmetric strongly centred monopoles.

The transformations of $(\eta,\zeta)$
corresponding to real $O(3)$ transformations of space can be calculated. For example,
rotation through an angle  $\phi$ around the $x_3$-axis is given by
\be
 R_\phi:(\eta,\zeta)\rightarrow(e^{i\phi}\eta,e^{i\phi}\zeta),
\ee
and inversion, $I$, is given by
\be
I:(\eta,\zeta)\rightarrow(\frac{-\bar{\eta}\,}
{\bar{\zeta}^2},\frac{1}{\bar{\zeta}}).
\ee
General $SO(3)$ transformations will correspond to M\"{o}bius
transformations of $\zeta$ which may be calculated. This allows us to
impose symmetry on candidate algebraic curves. Strong centring can
easily be imposed \cite{HMM} on the spectral curve. For strongly
centred
 monopoles
\be a_1(\zeta)=0. \ee
We can now construct the candidate algebraic curve. It is
\be \eta^3+f\eta\zeta^2+ig(\zeta^5-\zeta)=0 \label{candac}\ee
where the constant coefficients $f$ and $g$ are real. The reality of
$g$ corresponds to choosing an orientation around the $x_3$-axis,
analogous to choosing $b$ real in (\ref{ratmap}).   

To show that there is a one-parameter set of $f$ and $g$ values such
that this algebraic curve is the spectral curve of a monopole we
employ the ADHMN construction. This is an alternative approach to the construction of
monopoles which nicely complements the \spc formulation. The ADHMN construction
\cite{N,HB} is an equivalence between $k$-monopoles and Nahm data
$(T_1,T_2,T_3)$, which are three $k\times k$ matrices depending
on a real parameter $s\in[0,2]$ and satisfying the following:\\

\newcounter{con}
\setcounter{con}{1}
(\roman{con})  Nahm's equation
\be
\frac{dT_i}{ds}=\half\epsilon_{ijk}[T_j,T_k], \label{Neqn}
\ee\\

\addtocounter{con}{1}
(\roman{con}) $T_i(s)$ is regular for $s\in(0,2)$ and has simple
poles at $s=0$ and $s=2$,\\

\addtocounter{con}{1}
(\roman{con}) the matrix residues of $(T_1,T_2,T_3)$ at each
pole form the irreducible $k$-dimensional representation of $SU(2)$,\\

\addtocounter{con}{1}
(\roman{con}) $T_i(s)=-T_i^\dagger(s)$,\\

\addtocounter{con}{1}
(\roman{con}) $T_i(s)=T_i^t(2-s)$.\\

\setcounter{con}{1}
Equation (\ref{Neqn}) is equivalent to a Lax pair. This gives an associated
algebraic curve which is given in terms of the Nahm data as 
\be
\mbox{det}(\eta+(T_1+iT_2)-2iT_3\zeta+(T_1-iT_2)\zeta^2)=0.
\label{curve}
\ee
By differentiating and substituting from (\ref{Neqn}) we
can see that this curve is independent of $s$. This algebraic
curve is, in fact, the spectral curve.

We now construct Nahm data invariant under the $I_4$ transformation. 
The Nahm data are an \R$^3\otimes sl(k,\C)$ valued function of $s$, which
transform under the rotation group $SO(3)$ as 
\begin{eqnarray}\underline{3}\otimes sl(\underline{k})
 &\cong&\underline{3}\otimes 
(\underline{2k-1}\oplus\underline{2k-3}\oplus
 \ldots \oplus \underline{3})\nonumber\\
&\cong&( \underline{2k+1}\oplus \underline{2k-1}\oplus
\underline{2k-3})\oplus \ldots \oplus (
\underline{5}\oplus \underline{3}\oplus\underline{1}).
\label{decomp}
\end{eqnarray}
where $\underline{r}$ denotes the unique irreducible $r$-dimensional
representation of $su(2)$. Thus for charge three monopoles the Nahm
data transform as the representation 
\be
(\underline{7}_u \oplus \underline{5}_m \oplus \underline{3}_l) \oplus
(\underline{5}_u \oplus \underline{3}_m \oplus \underline{1}_l)
\ee
where the subscripts $u, m$ and $l$, standing for upper, middle and lower, are a notational convenience to
allow us to distinguish between isomorphic representations with
different pedigrees.
The most convenient way of finding $I_4$-invariant vectors in these
representations is to represent $su(2)$ on homogeneous polynomials
over \CP$^1$. Write $X,Y$ and $H$ for the basis of $SU(2)$ satisfying
the 
commutation relations
\be [X,Y]=H;\qquad [H,X]=2X;\qquad [H,Y]=-2Y.\ee 
The $(r+1)$-dimensional $su(2)$ representation $\underline{r+1}$ is
defined
on degree $r$ homogeneous polynomials by the identification 
\be X=\zeta_1 \frac{\partial}{\partial \zeta_0};\qquad Y=\zeta_0
\frac {\partial}{\partial \zeta_1};\qquad H=-\zeta_0
\frac{\partial}{\partial \zeta_0}+\zeta_1 \frac{\partial}{\partial
\zeta_1}.\ee

It is easy to see that the degree four polynomial 
\be 
p_4(\zeta_1,\zeta_0)=\zeta_0^2\zeta_1^2
\ee
is invariant under $U(1)$ transformations around the $x_3$-axis. It
has two of its zeros on each of the two poles of the Riemann
sphere. The 
degree four polynomial 
\be 
q_4(\zeta_1,\zeta_0)=\zeta_1^4+\zeta_0^4
\ee
is invariant under $D_4$
transformations; its zeros are arranged with $C_4$ symmetry around the
equator of the Riemann sphere. Furthermore, up to a choice of
orientation, the one-parameter family of homogeneous polynomials 
\be \zeta_1^4+ih\zeta_1^2\zeta_0^2+\zeta_0^4 \ee
constitute all the $I_4$-invariant vectors in $\underline{5}$.
Similarly  
\be p_6(\zeta_1,\zeta_0)=\zeta_1^3\zeta_0^3 \ee
and
\be q_6(\zeta_1,\zeta_0)=\zeta_1\zeta_0(\zeta_1^4-\zeta_0^4) \ee
are degree six polynomials invariant under $I_4$ and
\be p_2(\zeta_1,\zeta_0)=\zeta_1\zeta_0 \ee
is a degree two $U(1)$ invariant.

Not all of these polynomials correspond to inhomogeneous polynomial
coefficients in the candidate algebraic curve (\ref{candac}). The polynomial $p_2$ is absent since strongly
centring the monopole forces the coefficient of $\eta^2$ to vanish. Furthermore not all
invariant polynomials over \CP$^1$ lift to invariant polynomials
over the entire tangent bundle.

We now construct a set of invariant Nahm triplets with closed commutation relations and
corresponding to the candidate spectral curve (\ref{candac}) and thus
to an $I_4$-symmetric 3-monopole. Such a set     
can be found by constructing
the triplets corresponding to the $D_4$-invariant vectors in
$\underline{7}_u$ and $\underline{5}_m$, the $U(1)$ invariant in
$\underline{5}_u$ and the $SO(3)$ invariant $\underline{1}_l$.
  
We use the usual scheme \cite{HMM,HSa,HSb} to construct Nahm triplets
corresponding to the given homogeneous polynomials. Polarizing $q_6$
gives
\begin{eqnarray}
  \xi_1^2\otimes 20\zeta_1^3\zeta_0+2\xi_1\xi_0\otimes
 5(\zeta_1^4-\zeta_0^4)+\xi_0^2 \otimes
  20\zeta_1\zeta_0^3\qquad\qquad\qquad\qquad\qquad\qquad\qquad\\ \sim(5\left( \zeta_0\frac{\partial}{\partial\zeta_1}\right)\zeta_1^4,5\zeta_1^4-\frac{5}{24}\left( \zeta_0 \frac{\partial}{\partial\zeta_1}\right)^4\zeta_1^4,\frac{5}{6}\left( \zeta_0\frac{\partial}{\partial\zeta_1}\right)^3\zeta_1^4).\nonumber\end{eqnarray}
We make the identification 
\be \zeta_1^{2r}\leftrightarrow X^r;\qquad\qquad\zeta_0\frac{\partial}{\partial\zeta_1}\leftrightarrow\mbox{ad}Y\ee 
and derive the Nahm triplet
\be (5 \mbox{ad} Y X^2 ,5 X^2-\frac{5}{24}\mbox{ad} Y^4 X^2 , \frac{5}{6} \mbox{ad} Y^3 X^2 ) .\label{derived}\ee
We choose the explicit basis 
\be 
X=  \left[ 
{\begin{array}{ccr}
0 & 0 & 0 \\
 - i\,\sqrt {2} & 0 & 0 \\
0 &  - i\,\sqrt {2} & 0
\end{array}}
 \right];\;\; 
Y=  \left[ 
{\begin{array}{rcc}
0 & i\,\sqrt {2} & 0 \\
0 & 0 & i\,\sqrt {2} \\
0 & 0 & 0
\end{array}}
 \right];\;\; H=  \left[ 
{\begin{array}{rrr}
-2 & 0 & 0 \\
0 & 0 & 0 \\
0 & 0 & 2
\end{array}}
 \right],\label{bbasis}
\ee
and substitute (\ref{bbasis}) into (\ref{derived}). For convenience we perform the 
 Nahm isospace basis transformation 
\be(S_1,S_2,S_3)=(\half S_1^\prime+S_3^\prime,-\frac{i}{2}
S_1^\prime+iS_3^\prime,-iS_2^\prime).\label{basis}\ee
Relative to this basis the $SO(3)$-invariant Nahm triplet
corresponding to the $\underline{1}$ representation in (\ref{decomp})
is given by $(\rho_1,\rho_2,\rho_3)$ where 
\be
\rho_1=X-Y; \qquad \rho_2=i(X+Y); \qquad \rho_3=iH,
\ee
and the $\underline{7}_u$ $D_4$-invariant Nahm triplet (\ref{derived}) is
\be 
W_1=  \left[ 
{\begin{array}{ccc}
0 & \sqrt {2} & 0 \\
 - \sqrt {2} & 0 &  - \sqrt {2} \\
0 & \sqrt {2} & 0
\end{array}}
 \right];\;\;
W_2 =  \left[ 
{\begin{array}{ccc}
0 & i\,\sqrt {2} & 0 \\
i\,\sqrt {2} & 0 &  - i\,\sqrt {2} \\
0 &  - i\,\sqrt {2} & 0
\end{array}}
 \right];\;\;
W_3=  \left[ 
{\begin{array}{rrr}
0 & 0 & 2 \\
0 & 0 & 0 \\
-2 & 0 & 0
\end{array}}
 \right].
\ee

The $U(1)$-invariant polynomial $p_4$ polarizes as
\begin{eqnarray} \xi_1^2\otimes 2\zeta_0^2+2\xi_1\xi_0\otimes
 4\zeta_1\zeta_0+\xi_0^2 \otimes
  2\zeta_1^2&\sim&(\left(
  \zeta_0\frac{\partial}{\partial\zeta_1}\right)^2\zeta_1^2,2\left(
  \zeta_0
  \frac{\partial}{\partial\zeta_1}\right)\zeta_1^2,\zeta_1^2)\nonumber\\
&\leftrightarrow&(\mbox{ad}Y^2X,2\mbox{ad}YX,X).\nonumber\end{eqnarray}
After performing the transformation (\ref{basis}) the $\underline{5}_u$ $U(1)$-invariant Nahm triplet is
\be 
Y_1=  \left[ 
{\begin{array}{ccc}
0 &  - i\,\sqrt {2} & 0 \\
 - i\,\sqrt {2} & 0 &  - i\,\sqrt {2} \\
0 &  - i\,\sqrt {2} & 0
\end{array}}
 \right];\;\;
Y_2 =  \left[ 
{\begin{array}{ccc}
0 &  - \sqrt {2} & 0 \\
\sqrt {2} & 0 &  - \sqrt {2} \\
0 & \sqrt {2} & 0
\end{array}}
 \right];\;\;
Y_3 =  \left[ 
{\begin{array}{crc}
4\,i & 0 & 0 \\
0 & 0 & 0 \\
0 & 0 &  - 4\,i
\end{array}}
 \right]. 
\ee
To construct the $D_4$ invariant in $\underline{5}_m$ we first construct the
invariant in $\underline{5}_u$. Polarizing $q_4$ gives, to a constant multiple,
 \begin{eqnarray} \xi_1^2\otimes \zeta_1^2+\xi_0^2 \otimes
  \zeta_0^4&\sim&\xi_1^2\otimes\zeta_1^4+\frac{1}{4}\left(
  \xi_0\frac{\partial}{\partial\xi_1}\right)^2\xi_1^2\otimes\left(
  \zeta_0
  \frac{\partial}{\partial\zeta_1}\right)^2\zeta_1^2\nonumber\\
&=&\left[1+\frac{1}{24}\left(
  \xi_0\frac{\partial}{\partial\xi_1}\otimes 1+1\otimes
  \zeta_0
  \frac{\partial}{\partial\zeta_1}\right)^4\right]\xi_1^2\otimes\zeta_1^2.\end{eqnarray}
We now map this invariant into $\underline{5}_m$ by replacing the $\underline{5}_u$ highest weight
vector $\xi_1^2\otimes\zeta_1^2$ with the
 $\underline{5}_m$ highest weight vector
\be \xi_1^2\otimes\zeta_0\zeta_1^3-\xi_0\xi_1\otimes\zeta_1^4.\ee
Expanding this we derive
$$ \xi_1^2\otimes 
\left(-6\left(\zeta_0\frac{\partial}{\partial\zeta_1}\right)
\zeta_1^4\right)+2\xi_1\xi_0\otimes
 \left(12\zeta_1^4+\frac{1}{2}\left(\zeta_0\frac{\partial}
{\partial\zeta_1}\right)^4\zeta_1^4\right)+\xi_0^2 
\otimes\left(-\left(\zeta_0\frac{\partial}{\partial\zeta_1}\right)
\zeta_1^4\right)$$
\begin{eqnarray}&\sim&(-6\left(\zeta_0\frac{\partial}{\partial\zeta_1}
\right)\zeta_1^4,12\zeta_1^4+\frac{1}{2}\left(\zeta_0\frac{\partial}
{\partial\zeta_1}\right)^4\zeta_1^4,-\frac{1}{2}
\left(\zeta_0\frac{\partial}{\partial\zeta_1}\right)^3\zeta_1^4)\nonumber\\
&\leftrightarrow&(-6\mbox{ad}YX^2,[12+\frac{1}{2}\mbox{ad}Y^4]X^2,
-\frac{1}{2}\mbox{ad}Y^3X^2).\end{eqnarray}
After performing the transformation (\ref{basis}) the $\underline{5}_m$ $D_4$-invariant Nahm triplet is
\be 
Z_1=  \left[ 
{\begin{array}{ccc}
0 &  - \sqrt {2} & 0 \\
\sqrt {2} & 0 & \sqrt {2} \\
0 &  - \sqrt {2} & 0
\end{array}}
 \right];\;\; 
Z_2=  \left[ 
{\begin{array}{ccc}
0 &  - i\,\sqrt {2} & 0 \\
 - i\,\sqrt {2} & 0 & i\,\sqrt {2} \\
0 & i\,\sqrt {2} & 0
\end{array}}
 \right];\;\;
Z_3=  \left[ 
{\begin{array}{rrr}
0 & 0 & 4 \\
0 & 0 & 0 \\
-4 & 0 & 0
\end{array}}
 \right].
\ee

The Nahm data are
\be
T_i(s)=x(s)\rho_i+y(s)Y_i+z(s)Z_i+w(s)W_i.
\ee
They give monopoles with the required symmetry.
We now know the matrices $\rho_i$, $Y_i$, $Z_i$ and $W_i$ and can
substitute $T_i(s)$ into Nahm's equation (\ref{Neqn}) so that it
reduces to equations for $x$, $y$, $z$ and
$w$. 

To solve Nahm's equation it is convenient to replace the variables 
$x,y,z,w$ by the new variables $\alpha,\beta,\theta,\phi$ defined via
\bea
3x&=&\alpha\cos\phi+2\beta\cos\theta \nonumber \\
3y&=&-\alpha\cos\phi+\beta\cos\theta \nonumber \\
3z&=&\alpha\sin\phi+\beta\sin\theta\\
3w&=&\alpha\sin\phi-2\beta\sin\theta. \nonumber
\eea
Then the Nahm data take the simple form
$$
T_1=i\beta\sqrt{2}\left[\begin{array}{ccc}
0&e^{-i\theta}&0\\
e^{i\theta}&0&e^{i\theta}\\
0&e^{-i\theta}&0
\end{array}
\right];
\;\; 
T_2=\beta\sqrt{2}\left[\begin{array}{ccc}
0&e^{i\theta}&0\\
-e^{-i\theta}&0&e^{-i\theta}\\
0&-e^{i\theta}&0
\end{array}
\right];
$$
\be
T_3=2\alpha\left[\begin{array}{ccc}
i\cos\phi&0&-\sin\phi\\
0&0&0\\
\sin\phi&0&-i\cos\phi
\end{array}
\right].
\ee

With these data, Nahm's equation becomes
\bea 
\dot\alpha &=& -2\beta^2\cos(2\theta-\phi) \nonumber\\
\dot\beta&=& -2\alpha\beta\cos(2\theta-\phi) \nonumber\\
\dot\theta&= & \ 2\alpha\sin(2\theta-\phi) \\ 
\dot\phi&= & -2\beta^2\alpha^{-1}\sin(2\theta-\phi). \nonumber
\eea
The spectral curve, calculated using (\ref{curve}),
is of the required form (\ref{candac}) with the constants $f$ and $g$
given by
\be
f=16(\beta^2-\alpha^2)\hskip 15pt , \hskip 15pt
g=32\alpha\beta^2\sin(2\theta-\phi).
\label{consta}
\ee

It is convenient, although at first it may appear a strange choice,
to trade-in the variables $f$ and $g$ for two new variables
$a$ and $\kappa$ defined by
\be
f=-2\kappa^2(a^2+4\epsilon)^{1/3} \hskip 15pt,\hskip 15pt
g=2\kappa^3a
\ee
where $\epsilon=\pm 1$ allows two choices for this transformation.
Using these we solve the equations 
for $\alpha$ and $\beta$ as
\bea
\alpha(s)&=&\frac{\kappa}{2}\sqrt{\wp(\kappa s)+(a^2+4\epsilon)^{1/3}}
\label{sola}
\\
\beta(s)&=&\frac{\kappa}{2}\sqrt{\wp(\kappa s)-\frac{1}{2}
(a^2+4\epsilon)^{1/3}} 
\eea
where $\wp$ is the Weierstrass elliptic function satisfying
\be
{\wp^\prime}^2=4\wp^3-3(a^2+4\epsilon)^{2/3}\wp+4\epsilon
\label{wpfun}
\ee
and prime denotes differentiation with respect to the argument. The
equation for $\phi$  then becomes
\be
\dot\phi=\frac{-\kappa a}{2(\wp+(a^2+4\epsilon)^{1/3})}
\label{phieq}
\ee
and $\theta$ is related to $\phi$ via
\be
\sin(2\theta-\phi)=\frac{\kappa^3 a}{16\alpha\beta^2}.
\ee

Let us now check that the above data satisfy the Nahm boundary
conditions and thus correspond to a monopole. For $\alpha$ and $\beta$ to be finite for $s\in(0,2)$
and to have simple poles at $s=0$ and $s=2$ requires that $\kappa$
is half the real period of the elliptic function given by
(\ref{wpfun}). This fixes the value of $\kappa$ for given $a$ and $\epsilon$.
It can easily be checked from the above equations that both
$\theta$ and $\phi$ are finite for $s\in[0,2]$.
Next we need to check that  $R_i$, the matrix residues of the poles of $T_i$, form the irreducible representation of $SU(2)$.
As $s\rightarrow 0$ then, by (\ref{sola}), 
\be
\alpha\sim \frac{1}{2s}
\ee
so that $iR_3$ has eigenvalues $\{0,\pm1\}$, independently of the
value of $\phi$ at $s=0$. This demonstrates that the representation formed
by the matrix residues is the irreducible one. A similar argument applies
for the pole at $s=2$. We have proved that the above 
Nahm data correspond to a monopole. Conveniently this is done without
having to explicitly solve (\ref{phieq}) which would require
performing an elliptic integral of the third kind.

In summary, we have shown that
\be
\eta^3-6(a^2+4\epsilon)^{1/3}\kappa^2\eta\zeta^2
+2i\kappa^3a(\zeta^5-\zeta)=0
\label{spec}
\ee
is the spectral curve of a 3-monopole with $I_4$ symmetry for all $a\in$\R\ and
 $\epsilon=\pm1$, provided $2\kappa$ is the real period of the elliptic
curve
\be
y^2=4x^3-3(a^2+4\epsilon)^{2/3}x+4\epsilon.
\label{ellipc}
\ee
We shall now analyse this family of spectral curves and examine
some particular cases in detail. For this we shall need the following
integral representation for $\kappa$,
\be
\kappa=\int_{x_0}^\infty \frac{dx}
{\sqrt{4x^3-3(a^2+4\epsilon)^{2/3}x+4\epsilon}}
\label{intrep}
\ee
where $x_0$ is the largest real root of
$4x^3-3(a^2+4\epsilon)^{2/3}x+4\epsilon=0$.

First we shall consider the case $\epsilon=-1$. Then $\kappa$ is
finite for all finite $a$. If $a=0$ then the Weierstrass function
given by (\ref{wpfun}) is degenerate, since there are two equal
real zeros of the elliptic curve (\ref{ellipc}). This is the case of
infinite imaginary period and in this case $\wp$ may be written
in terms of trigonometric functions as
\be
\wp(z)=2^{-1/3}(\frac{3}{\sin^2(2^{-1/6}\sqrt{3}\/z)}-1).
\ee
The real period of this function must be $2\kappa$,  hence
\be
\kappa=\frac{\pi}{2^{5/6}\sqrt{3}}.
\ee
Substituting these values into (\ref{spec}) gives the curve
\be
\eta^3+\pi^2\eta\zeta^2=0
\ee
which is the spectral curve of the axisymmetric 3-monopole
\cite{HA}.

If $a=\pm2$ then 
\be
\kappa=\int_1^\infty\frac{dx}{2\sqrt{x^3-1}}
=\frac{\Gamma(1/6)\Gamma(1/3)}{4\sqrt{3\pi}}
\ee
and (\ref{spec}) becomes
\be
\eta^3\pm i
\frac{\Gamma(1/6)^3\Gamma(1/3)^3}
{48\sqrt{3}\pi^{3/2}}
(\zeta^5-\zeta)=0 \label{spectet}
\ee
which is the spectral curve of the  
tetrahedral monopole \cite{HMM} in two different orientations.

Now consider the limit $a\rightarrow\infty$. An asymptotic
expansion of the integral representation (\ref{intrep}) gives
\be
\kappa\sim\frac{\Gamma(1/4)^2}{4a^{1/3}3^{1/4}\sqrt{\pi}}
\ee
so that (\ref{spec}) tends to
the limiting spectral curve
\be
\eta^3-\frac{\Gamma(1/4)^4\sqrt{3}}
{8\pi}\eta\zeta^2+i\frac{\Gamma(1/4)^6}{32\pi^{3/2}3^{3/4}}(\zeta^5-\zeta)=0.
\label{speceight}
\ee
This is a new explicit spectral curve which describes three monopoles
which are not well-separated. We shall see later that the energy
density of this monopole configuration has a complicated and unusual
structure. It resembles a twisted figure-of-eight, with
the bottom loop at right angles to the top loop.
The same monopole configuration is obtained in the limit 
$a\rightarrow-\infty$, but this time rotated by $90^\circ$ around the
$x_3$-axis.

For $\epsilon=-1$ we have seen that, although the parameter $a$ 
ranges over the whole real line, this maps out only a finite segment in the 3-monopole moduli space {\sl ie} there are
no solutions representing well-separated monopoles for $\epsilon=-1$.
>From the rational map approach of Section 2 we know that a
one-parameter family of solutions exists which includes 
well-separated monopoles. This implies that another branch of 
solutions exists which continues the $\epsilon=-1$ branch discussed
so far. This is indeed true and is given by setting $\epsilon=1$.
Let us now consider this case.

It is immediately clear from the general spectral curve
(\ref{spec}) and the integral representation (\ref{intrep}) that
the values
\be
\epsilon=-1, a\rightarrow\infty \mbox{ and } 
\epsilon=1, a\rightarrow\infty
\ee
describe the same spectral curve and hence the same monopole 
solution. This is the twisted figure-of-eight solution which
describes three monopoles close together. The same is true for $a\rightarrow-\infty$ with $\epsilon=\pm 1$.

Next consider the limit $a\rightarrow 0$, with $\epsilon=1$.
In this limit the elliptic curve (\ref{ellipc}) has a double
zero at the largest positive real root. Hence from (\ref{intrep}) 
$\kappa$ tends to infinity logarithmically with $a$ in this limit.
The spectral curve (\ref{spec}) is then asymptotic to 
\be
\eta^3-4^{1/3}6\kappa^2\eta\zeta^2=0.
\label{prodstar}
\ee
The product of three spectral curves describing unit charge monopoles
with centres $(x_1,x_2,x_3)$ given by
\be
\{(0,0,0),(0,0,-c),(0,0,+c)\}
\ee
is
\be
\eta^3-4c^2\eta\zeta^2=0.
\ee
The curve (\ref{prodstar}) has this form with
$c=2^{-1/6}\sqrt{3}\kappa$.

In summary, the full family of monopole solutions is described,
in our coordinates,
by a family of spectral curves which has three connected segments
given by $\epsilon=1,a>0;\ \epsilon=-1,a\in\R;\  \epsilon=1,a<0.$

The Higgs field $\Phi$ can be reconstructed from the Nahm data using the
ADHMN algorithm. This we have done using a numerical implementation of the ADHMN algorithm  introduced by the authors in
a previous paper \cite{HSa}. The energy density can then
be calculated by using the formula (\ref{wardsform}). The numerical algorithm requires the Nahm data at $s$-values on a
grid. Since we do not have a convenient analytic expression for $\phi$ the
equation (\ref{phieq}) is solved numerically, but this is a
simple task.

\begin{center}
\begin{tabular}{|c|c|c|} \hline
image & $\epsilon$ & a \\
\hline
1 & +1 & 0.300 \\
2 & +1 & 0.600 \\
3 & +1 & 1.000 \\
4 & -1 & 4.000 \\
5 & -1 & 2.010 \\
6 & -1 & 2.000 \\
7 & -1 & 1.995 \\
8 & -1 & 1.800 \\
9 & -1 & 0.000 \\
10 & -1 & -1.800 \\
11 & -1 & -1.995 \\
12 & -1 & -2.000  \\
13 & -1 & -2.010 \\
14 & -1 & -4.000 \\
15 & +1 & -1.000 \\
16 & +1 & -0.600 \\
17 & +1 & -0.300 \\
\hline
\end{tabular}
\end{center}
\begin{center}
Table 1. {\sl Parameter values for scattering shown in Fig. 1}
\end{center}\ \\

Fig. 1\footnote{Fig 1. is not available in the
hep-th version of this paper.
A hard copy is available on request to P.M.Sutcliffe@ukc.ac.uk,
or it can be viewed at URL
http://www.ukc.ac.uk/IMS/maths/people/P.M.Sutcliffe/preprints.html} 
 displays a surface of constant energy density
${\cal E}=0.18$  for 17 different members of the family of monopoles. Table 1 gives the values of the parameters
$\epsilon$ and $a$ for each image.
Since these monopole configurations all lie on the generator of the
surface $\Sigma_3^2$ we can now describe the monopole scattering
corresponding to motion along that generator in more detail than before.
At large negative times (1) there are
three well-separated monopoles.
One monopole is stationary at the origin, a second 
monopole is approaching along the positive $x_3$-axis and a third is
approaching along the negative $x_3$-axis. As the monopoles merge (2) the
one in the centre twists as it attempts to align with both the top and
bottom monopoles.
The energy tries to flow towards the centre but gets squeezed out 
sideways to form the twisted figure-of-eight shape (4). This is the
configuration given by the explicit spectral curve (\ref{speceight}).
The energy continues to flow towards the $x_1x_2$-plane, but
now it has more of a sideways motion, which leads to the formation
of the tetrahedral monopole (6). The diagonal movement
of the energy density pulls the tetrahedron apart (7) into a buckled
torus (8), which then straightens out to form the axisymmetric 
charge three torus (9) at time zero. The energy continues to flow in the same direction
so that the torus buckles in the opposite sense (10). The motion at positive
times goes backwards through the configurations
just described, except that the monopoles are inverted so that
the tetrahedral monopole (12) formed at positive time
is dual to the one (6) formed at negative time etc.\\

\section{The zeros of the Higgs field}
\news
\ \indent 
The monopole dynamics which occurs here is very novel 
and unlike any previously known.
In this section we shall discuss a surprising aspect of the charge
three twisted line scattering process. It appears that during this
scattering process the number of zeros of the Higgs field is not
conserved. The total number of zeros of the Higgs field {\sl counted with
their multiplicity} is $k$ for a $k$-monopole and these zeros are not
always isolated, but may coalesce to form
zeros of higher multiplicity. For example, in the case of the
 toroidal
3-monopole there is a single zero but it has multiplicity three. What
is surprising about the charge
three twisted line scattering geodesic is that there are intervals
when the total
number of zeros exceeds three.

The scattering process passes from three well-separated monopoles through
the tetrahedral configuration to the toroidal configuration. When there
are three well-separated monopoles the Higgs field has exactly three zeros.
The
axisymmetric monopole has all three Higgs zeros at
the origin and it is clear that the only way to
arrange three points with tetrahedral symmetry is to
put all three points at the origin. Thus if the 
tetrahedral monopole has three zeros of the Higgs field
then they must all be at the origin, as in the case
of the axisymmetric monopole. Moreover throughout twisted line
scattering the imposed symmetry means that if there are three zeros of the Higgs field
then one must be at the origin with the other two
on the $x_3$-axis and equidistant from the origin.
However, numerical investigations reveal that there
are no zeros of the Higgs field on the $x_3$-axis
(except at the origin) for all the monopole solutions
between the tetrahedral monopole (Fig 1 (5)) and the
axisymmetric monopole (Fig 1 (9)). So, if the number of
Higgs zeros remains three, then we are forced to the
surprising conclusion that the zeros of the Higgs field
must stick at the origin for a finite period of time.
It turns out that this is not in fact what is happening,
and the true description is even more fascinating.

The above argument fails because it assumes that
the number of zeros of the Higgs field is always three
for a 3-monopole solution of the Bogomolny equation.
The basis for such an assumption is that no $k$-monopole
solution has been presented which had greater
than $k$ zeros of the Higgs field and
furthermore  in the analogous case of
abelian Higgs vortices at critical coupling the 
vortex number not only
gives the total number of zeros counted with their multiplicity but
also bounds the total number of zeros \cite{JT} pp 76-78. 
However, 
what we have 
found is that some of the monopoles in our 
one-parameter family have more than three zeros.\footnote{We are
  extremely
 grateful to Werner Nahm for
suggesting this possibility} In fact, at different points
on the one-parameter family the number of zeros can be
one, three, five or seven.

The first approach we take is to compute the winding
number, $Q(r_0)$, of the normalized Higgs field on a
two-sphere of radius $r_0$, centred at the origin.
This integer winding number counts the number of zeros of the Higgs
field counted with multiplicity 
inside this two-sphere. By definition $Q(\infty)=k$ for
a $k$-monopole. The numerical scheme used to compute
the winding number is described in the appendix. 
The results obtained are integer valued to within six
decimal places, so we shall give all our results as 
integers. 

First we consider the tetrahedral monopole and 
 compute that $Q(1.0)=+3$. This is a good check on
our numerical scheme as we require that the winding
number is equal to three when $r_0$ is sufficiently 
large. Now if all three Higgs zeros were at the
origin then the winding number would equal three
for all positive values of $r_0$. However we find the
result that $Q(0.2)=-1$ ie. locally around the
origin the field configuration is that of an 
anti-zero. Therefore between the sphere of
radius $r_0=0.2$ and the sphere of radius $r_0=1.0$
there must be (at least) four zeros, which each
have an associated local winding number of +1.
Let us now look for these extra zeros by plotting the
components of the Higgs field.
 Write the Higgs field in terms of Pauli
matrices as \be
\Phi=i\sigma_1\varphi_1+i\sigma_2\varphi_2
+i\sigma_3\varphi_3
\ee
and plot the individual components
 $\varphi_1,\varphi_2,\varphi_3$. It is easier
to locate a zero of the Higgs field by searching
for where all three components are zero rather than
simply looking at the single quantity $\|\Phi\|$.
The task is made simple by the presence of tetrahedral
symmetry. Fig 2(a) shows the components of the 
Higgs field along the line $x_1=x_2=x_3=L$, for
$-0.4\le L\le 0.4$. It is clear that all three
curves have a zero at $L=0$ and $L\approx -0.38$.
By tetrahedral symmetry, similar curves are obtained
along each of the other three diagonals. Hence the
numerical evidence suggests that there are four 
positive zeros (ie. each corresponding to a winding of $+1$)
on the vertices of a regular tetrahedron and an
anti-zero (ie. corresponding to a winding of $-1$) at the
origin. Therefore the tetrahedral 3-monopole is
 a solution in which the
Higgs field has both positive multiplicity and 
negative multiplicity
zeros but nonetheless saturates the Bogomolny
energy bound. 

An obvious question is whether the positive zeros
lie along the directions of the vertices of the
tetrahedron (where the energy density is maximal)
or along the directions of the faces of the tetrahedron
(where a surface of constant energy density has holes).
Fig 2(b) shows a plot of the energy density along
the line  $x_1=x_2=x_3=L$, for $-3\le L\le 3$.
Clearly the zeros lie along the lines joining the
origin to the vertices of the tetrahedron. However
the zeros are not as far from the origin as the
points of maximal energy density. The zeros
occur at $L \approx -0.38$ whereas the energy density
takes its maximum value at $L \approx -1$.
It is interesting to note that the location of
the anti-zero appears to coincide with a local 
minimum of the energy density.

We are now in a position to describe the 
motion of the zeros of the Higgs field. We have roughly sketched this
motion in Fig 3. When the monopoles are well separated there are three
zeros of the Higgs field, Fig 3(a). One is at the origin and the other
two are equidistant above and below the origin along the $x_3$-axis. Obviously
in the asymptotic limit of infinite separation each of these zeros
lies at the centre of a 1-monopole. At some critical point as the
zeros approach there is a bifurcation. Each of the zeros above and
below the origin split into three zeros, two with positive
multiplicity and one with negative multiplicity. In this way the
number of zeros counted with their multiplicity is conserved
locally. Unfortunately neither the precise details of this bifurcation
nor the precise point at which it occurs is discernable
 numerically but
it is certain that as the matter begins to coalesce there are seven zeros
of the Higgs field, Fig 3(b), one at the origin of positive unit multiplicity,
two above and below the origin on the $x_3$-axis of negative unit
multiplicity and four further positive multiplicity zeros away from
the $x_3$-axis. These four zeros move consistently with the twisted
line symmetry, the two above the $x_1x_2$-plane are separated along a
line parallel to the line
$x_1=-x_2$ and $x_3=0$ and the two below, parallel to the line
$x_1=x_2$ and $x_3=0$. 

As the monopoles continue to coalesce the anti-zeros
approach the origin, Fig 3(c). They leave behind the four zeros which
are off the $x_3$-axis. They reach the origin at the tetrahedral configuration, Fig. 3(d). At the tetrahedral
configuration there are five zeros, a single zero of negative unit
multiplicity at the origin and four with positive multiplicity
arranged in a tetrahedron. The four positive multiplicity zeros then move
towards the origin, Fig 3(e) and finally reach the origin to give a
single multiplicity three zero, Fig 3(f).

We have already presented numerical evidence to
support our claim for the configuration of Higgs zeros
at the parameter value $a=2$ (the tetrahedral monopole).
Now we shall give similar numerical results to support
our claims for the configuration of Higgs zeros prior
to the formation of the tetrahedral monopole, but
after the splitting of the Higgs zeros.

The case considered is for the parameter value $a=2.05$.
First we compute some winding numbers. The results are
that 
\be Q(0.2)=+1, \ Q(0.5)=-1, \ Q(0.7)=+3 \ee
These results are consistent with a positive zero
at the origin, two negative zeros on the $x_3$-axis
and four positive zeros which are further from the
origin than the negative zeros.
The positions of the zeros can be located, in the
same manner as in the tetrahedral case, by plotting
the components of the Higgs field. By the imposed
symmetries each of the zeros must lie on a line where $x_1=\pm x_2$.
Therefore we plot the components of the Higgs field
along the line $x_1=x_2=L$, with $x_3$ fixed.
Fig 4(a) shows such a plot with $x_3=-0.425$.
This clearly shows a zero on the $x_3$-axis (ie. $L=0$).
Fig 4(b) shows a similar plot for $x_3=-0.605$.
It can be seen that there are two zeros, which are
a distance $L\approx 0.17$ from the $x_3$-axis.
These results are in agreement with the winding number
calculations, which placed bounds on the distances
of each of the zeros from the origin. 

It is difficult to give a physical interpretation of 
these negative
multiplicity zeros of the Higgs field. 
A natural interpretation would be that the configuration
contains anti-monopoles as well as monopoles, but such
a statement can not be made rigorous unless a suitable
definition of magnetic charge density is found. The
standard definition of the magnetic field of a monopole
relies upon a consideration of the asymptotic field of
the configuration far from the monopole, where the
non-abelian $SU(2)$ gauge symmetry is broken to
a $U(1)$ symmetry that can be identified with
the abelian gauge symmetry of electromagnetism.
In the twisted line scattering configurations the negative
multiplicity zeros remain
close to the positive multiplicity zeros and so a definition of
magnetic field that is valid only in the asymptotic region is
useless.

The existence of anti-zeros raises a number of mathematical questions.
For example, can the presence
of an anti-zero be seen from the spectral curve
or rational map of a monopole? It seems likely that the appearance and
disappearance of anti-zeros has a signature in the space of rational
maps or in the space of spectral curves. Unfortunately our numerical
results cannot pin-point the  exact values of the parameter $a$
 at which
anti-zeros appear or disappear in the 3-monopole
twisted line scattering. One possibility is that
such an event is associated with the
elliptic curve (\ref{ellipc}) 
being singular. The discriminant $\Delta$ of the
elliptic curve is
\be
\Delta=27a^2(a^2+8\epsilon).
\ee
This vanishes when $a=0$ and when $a=\pm\sqrt{8}, \ 
\epsilon=-1$. The first singular curve corresponds to
the toroidal monopole and we know that 
anti-zeros disappear and appear at this point. It is
consistent with our numerical results that the second
singular curve $a=\sqrt{8}, \ \epsilon=-1$ (and its
inverted partner) corresponds to the splitting point
where anti-zeros appear (disappear). It is also interesting to note
that a naive parameter count demonstrates that the motion of the
Higgs zeros and anti-zeros cannot be independent.

\section{Conclusion}
\news
\ \indent
By imposing twisted inversion symmetries on multi-monopoles we
have been able to present some interesting new examples
of geodesic monopole scattering. These scattering events help
to explain how the recently discovered monopoles with the
symmetries of the regular solids are formed from individual
 monopoles
as they merge and deform. The scattering processes are quite 
exotic
and reveal new features which are not yet fully understood.

The discovery that there exist $k$-monopole solutions
with greater than $k$ zeros of the Higgs field is
unexpected.  It is also surprising to find a large family of
scattering events in which the scattering angle is zero.
It seems that the behaviour
of higher charge monopoles is not completely typified by that
of charge two monopoles and
the behaviour of three-dimensional solitons is not typified by that of
two-dimensional solitons.
It remains to be seen whether these results concerning monopole scattering
are relevant to the dynamics of other three-dimensional 
topological solitons such as skyrmions.

On a more speculative
note, it seems likely that the creation of extra Higgs zeros is
connected with points in the monopole 
moduli space related to singular elliptic curves.
Singular points of moduli spaces are relevant
to phase transitions in string theory and 
are a central feature of the
Seiberg-Witten \cite{SW}  treatment of duality in
$N=2$ supersymmetric Yang-Mills theory, where they
are associated with the appearance of massless monopoles.
Perhaps some insight into these issues may be obtained
from a deeper study of the BPS monopole moduli space
at points where anti-zeros occur.
\\

\noindent{\bf Acknowledgements}

Many thanks to Nigel Hitchin, Trevor Samols
and Paul Shah for useful 
discussions. We would like to thank Nick Manton for many discussions
and advice during the preparation of this paper. CJH thanks the EPSRC for a research studentship and the
British Council for a FCO award. PMS thanks the EPSRC
for a research fellowship.\\[3cm]

\newpage
\appendix
\section{Appendix: Computation of winding numbers}
\news
\renewcommand{\theequation}{A\arabic{equation}}
\ \indent 
In this appendix we give the details of the numerical scheme
used to compute the winding number $Q$, of the normalised
Higgs field on a two-sphere of radius $r_0$, centred
at the origin.

First of all, we discretise the above two-sphere into the
$n^2$ lattice points given by
\bea
x_1&=&r_0\sin(\pi i/n)\cos(2\pi j/n)\nonumber \\
x_2&=&r_0\sin(\pi i/n)\sin(2\pi j/n)\\
x_3&=&r_0\cos(\pi i/n)\nonumber 
\eea
with $i=0,1,..,n-1$ and $j=0,1,..,n-1$.
Then using the  numerical ADHMN construction
\cite{HSa} the Higgs field is computed at each of these
lattice points. Write the Higgs field in terms of Pauli
matrices as
\be
\Phi=i\sigma_1\varphi_1+i\sigma_2\varphi_2
+i\sigma_3\varphi_3.
\ee
Then, providing the Higgs field is not identically
zero, we can define the unit 3-vector $\psi$ as
\be
\psi=(\varphi_1,\varphi_2,\varphi_3)
\frac{1}{\sqrt{\varphi_1^2+\varphi_2^2+\varphi_3^2}}.
\ee
So there is a unit 3-vector defined at each lattice point
 on the two-sphere. Fig 5 shows four such lattice
points, which are numbered 1 to 4. They correspond
to the lattice points $(i,j),(i+1,j),(i,j-1),(i+1,j-1)$,
for some integers $i$ and $j$. Let the unit 3-vectors
defined at these points be denoted by $\psi_1,\psi_2,
\psi_3,\psi_4$ respectively.

Now we can make use of the work of Berg and L\"uscher
\cite{BL} who have defined a lattice topological
charge for the $O(3)$ $\sigma$-model.
Let $Q_{ij}$ be the lattice topological charge density
\be
Q_{ij}=(A_{123}+A_{134})/4\pi
\ee
where $A_{123}$ denotes the signed area of the
spherical triangle with vertices $\psi_1,\psi_2,\psi_3$.
Explicitly the formula is
\be
\exp(\frac{iA_{123}}{2})=
\frac{1+\psi_1\cdot\psi_2+\psi_2\cdot\psi_3
+\psi_3\cdot\psi_1+i\psi_1\cdot(\psi_2\times\psi_3)}
{\sqrt{2(1+\psi_1\cdot\psi_2)(1+\psi_2\cdot\psi_3)
(1+\psi_3\cdot\psi_1)}}.
\ee
$4\pi Q$ is defined to be the total signed area of the
 surface obtained by glueing together all these 
elementary spherical triangles ie.
\be
Q=\sum_{i=0}^{n-1}\sum_{j=1}^n Q_{ij}.
\ee
By the geometrical interpretation of $Q$ it is
clear that it is integer valued. Moreover,
 it is topological in the sense that local continuous
deformations of the lattice field $\psi$
do not change the winding number $Q$ (providing
certain exceptional configurations are excluded)
\cite{BL}.\\

\newpage
\noindent{\bf Figure captions}\\

Fig. 1: Surface of constant energy density ${\cal E}=0.18$ at
increasing times.\\

Fig. 2(a): Components of the Higgs field of the
tetrahedral monopole along the line $x_1=x_2=x_3=L$,
for $-0.4\le L\le 0.4$.\\

Fig. 2(b): Energy density of the tetrahedral monopole
along the line $x_1=x_2=x_3=L$,
for $-3\le L\le 3$.\\

Fig. 3: Schematic representation of the motion of
the zeros of the Higgs field.\\

Fig. 4(a): Components of the Higgs field of the
 monopole with parameter $a=2.05$, along the
line $x_3=-0.425$, $x_1=x_2=L$, for $-0.4\le L\le 0.4$.\\

Fig. 4(b): As Fig. 4(a) but with $x_3=-0.605$.\\

Fig. 5: A vector field on a lattice $S^2$.

\newpage
\begin{figure}[ht]
\begin{center}
{\Large \bf Fig. 2}
\end{center}
\vskip -3cm
\hskip 2cm a{\epsfxsize=8cm \epsffile{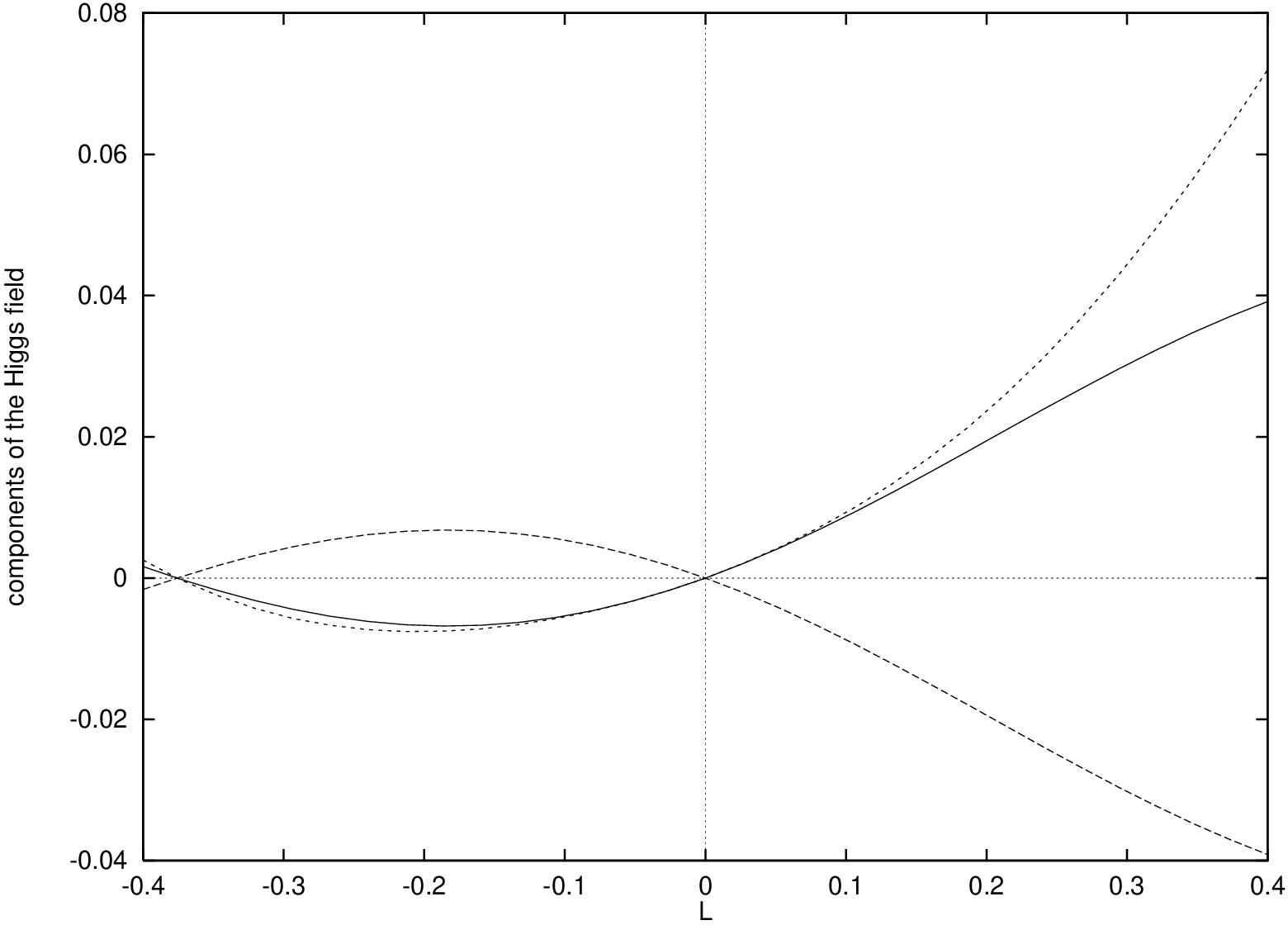}}
\vskip -3cm
\hskip 2cm b{\epsfxsize=8cm \epsffile{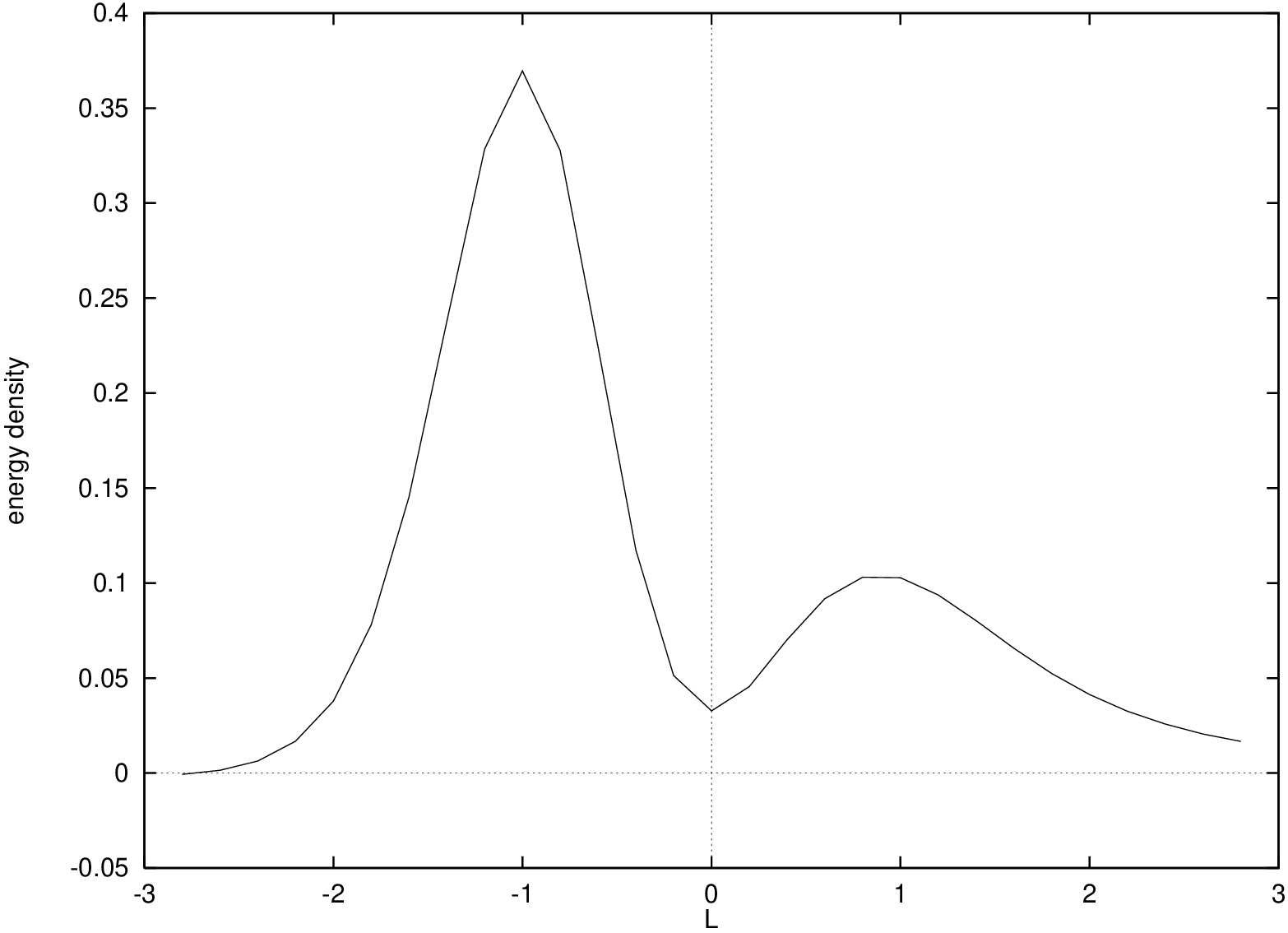}}
\end{figure}
\newpage

\begin{figure}[ht]
\begin{center}
\vskip 1cm
\epsfxsize=16cm \epsffile{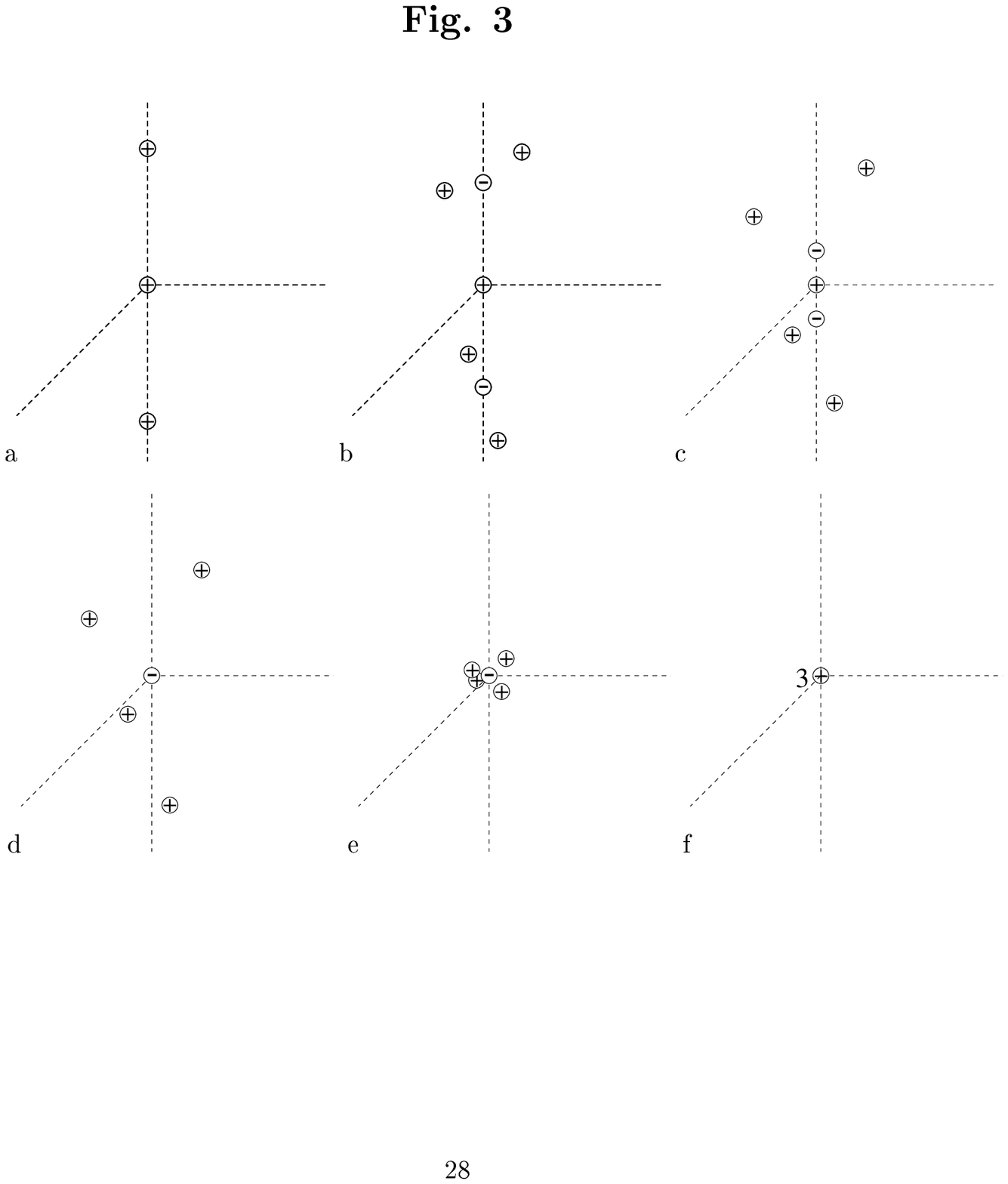}
\end{center}
\end{figure}

\begin{figure}[ht]
\begin{center}
{\Large \bf Fig. 4}
\end{center}
\vskip -3cm
\hskip 2cm a{\epsfxsize=8cm \epsffile{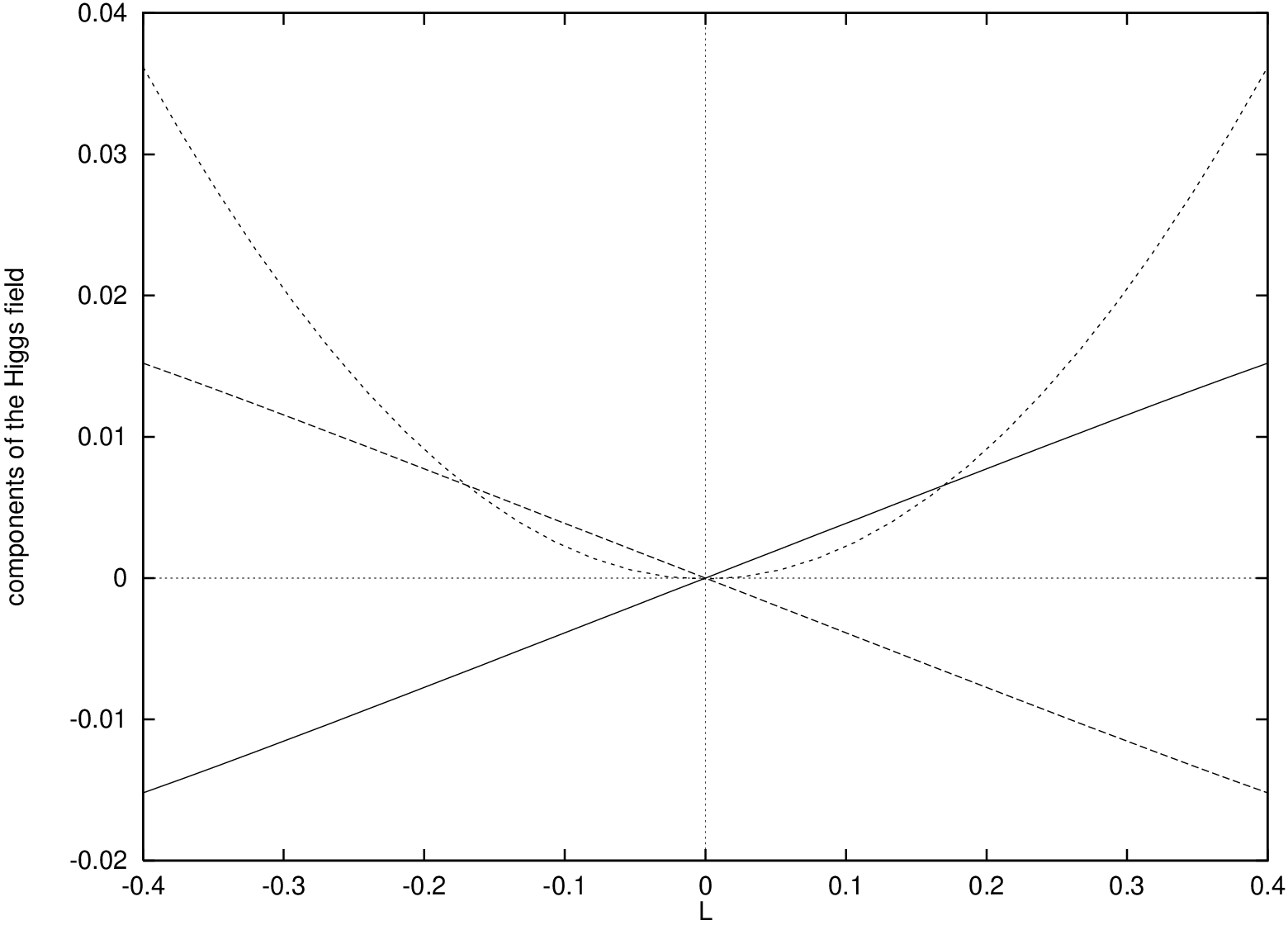}}
\vskip -3cm
\hskip 2cm b{\epsfxsize=8cm \epsffile{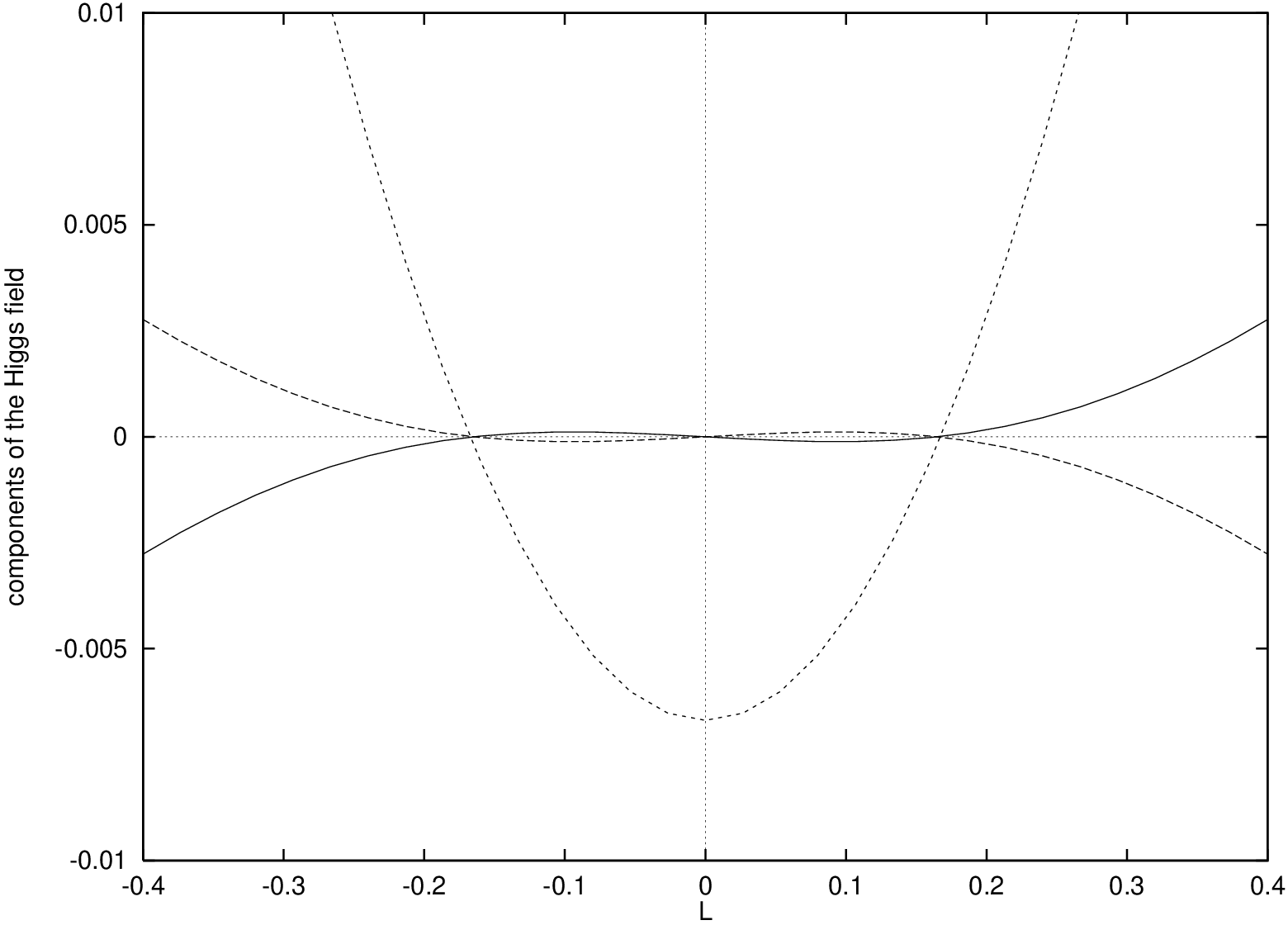}}
\end{figure}

\begin{figure}[ht]
\begin{center}
{\Large \bf Fig. 5}
\vskip 1cm
\leavevmode
{\epsfxsize=12cm \epsffile{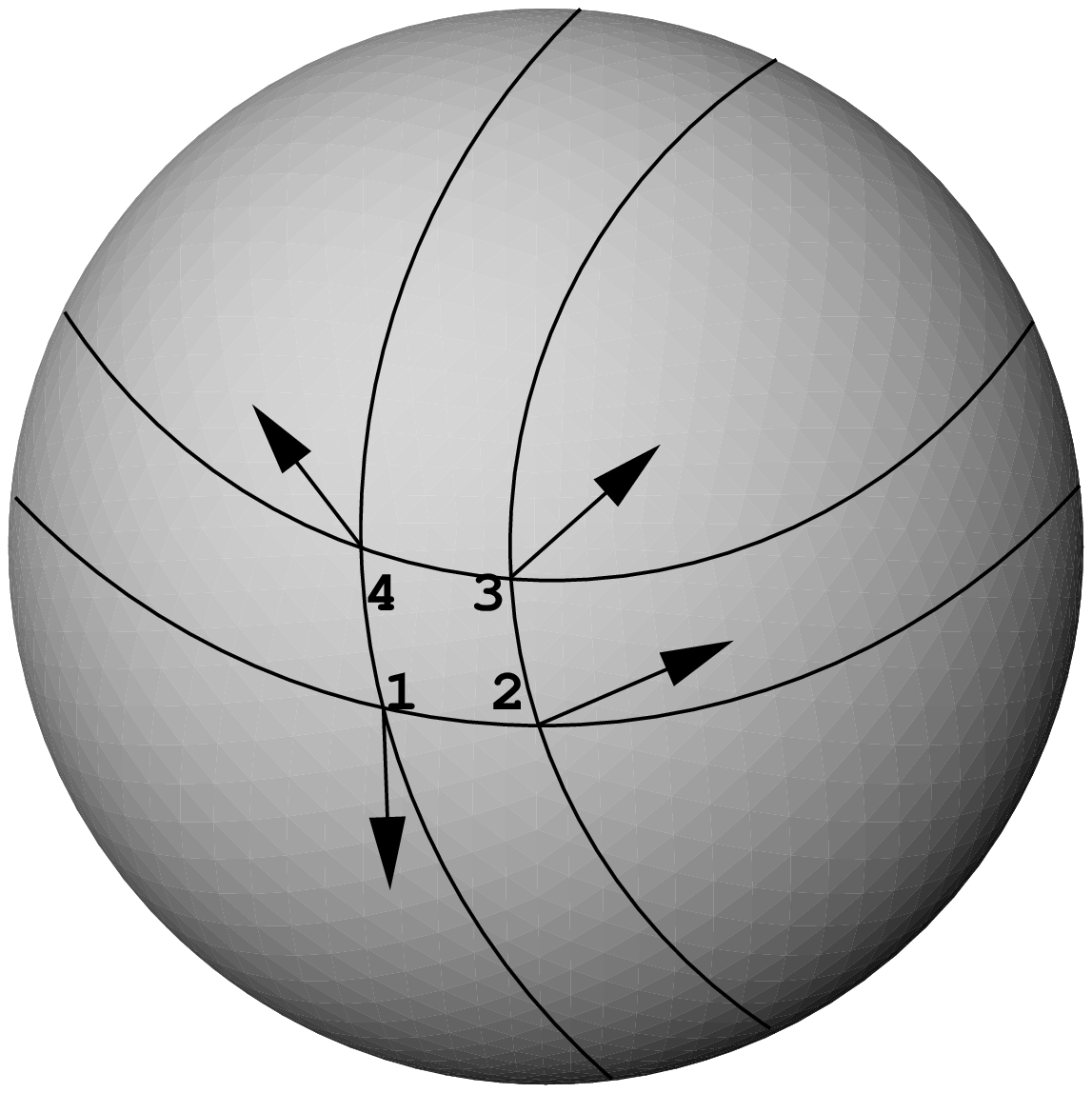}}
\end{center}
\end{figure}

\end{document}